\documentclass[prd,preprintnumbers,nofootinbib,aps,twocolumn]{revtex4-1}
\usepackage{epsfig,amsmath,amssymb}
\usepackage{mathrsfs}
\usepackage[normalem]{ulem}
\usepackage[usenames,dvipsnames]{color}
\usepackage[pagebackref=false, colorlinks=false]{hyperref}

\begin{document}
\title{Can Gravitational Wave Data Shed Light on Dark Matter Particles ? }

\author{Parthasarathi Majumdar}
\email{bhpartha@gmail.com}
\affiliation{School of Physical Sciences, Indian Association for the Cultivation of Science, Kolkata 700032, India.} 

\begin{abstract}

Gravitational wave (GW) data from observed binary black hole coalescences (BBHC) have been demonstrated in recent analyses to validate the Hawking Area Theorem (HAT) for black hole horizons. The result of such analyses is imposed here as a criterion of  {\it absolute} consistency on the logarithmic (in horizon area) corrections to the Bekenstein-Hawking Area Formula (BHAF) for the black hole entropy, when these corrections are computed both from non-perturbative quantum fluctuations of spacetime in matter-free quantum general relativity, as well as arising due to perturbative quantum matter field fluctuations around a stationary classical black hole background spacetime. This criterion of absolute consistency is seen to be obeyed provided certain restrictions ensue on the spin-parity and number of species of the spectrum of quantum matter fluctuations. Such constraints appear to restrict the Beyond-Standard-Model (BSM) part of the matter fluctuation spectrum. Some species of the constrained, yet-unobserved BSM particle spectrum are currently under active consideration in particle cosmology as candidates for dark matter.                   

\end{abstract}

\maketitle

\section{Introduction}

The issue of consistency of the {\it leading order} corrections to the Bekenstein-Hawking Area Formula (BHAF) \cite{bek73}-\cite{haw75} for black hole entropy, which are logarithmic in the horizon area, with gravitational wave (GW) observational results \cite{teu20}, \cite{bad22} validating Hawking's Area Theorem (HAT) \cite{haw71} for astrophysical binary black hole coalescences (BBHC), have been considered in two recent publications \cite{pm2024}, \cite{pm2025}. Such a consistency is demonstrated in \cite{pm2024} for logarithmic corrections calculated within a specific approach to quantum general relativity, especially quantum isolated black holes, namely matter-free pure Loop Quantum Gravity (LQG). This result uses earlier computations of logarithmic corrections to the BHAF within LQG \cite{km98}-\cite{abhi-pm14} . In ref. \cite{pm2025}, general criteria for {\it absolute} consistency of GW observations on BBHC, with the algebraic sign of general logarithmic corrections to the BHA formula are presented, irrespective of the methodology adopted for the computation of such corrections for black holes with macroscopically large horizon areas. The criterion of {\it absolute} consistency with the GW observational results is imposed on the logarithmic corrections, resulting in fixing the algebraic sign of the coefficient of the logarithmic corrections. 

The modifications to classical general relativity, arising due to quantum spacetime and matter fluctuations, is deemed \cite{pm2025} to be the source of logarithmic corrections to the BHAF for black hole entropy. Generically, for astrophysical black holes of macroscopically large horizon areas relative to the Planck area, the corrections to the BHAF may be thought of as a power series in inverse horizon area, with coefficients being functions of new `hair' arising from the quantization of spacetime  and matter. However, this power series must also include {\it logarithmic} (in horizon area) terms because if we consider the zeroth order term in the power series in a limiting sense : $\epsilon \rightarrow 0$ for $s_{\epsilon} S_{BH}^{\epsilon} = s_{\epsilon} \exp (\log S_{BH}^{\epsilon}) \simeq s_{\epsilon} (1 + \epsilon \log S_{BH})$ where $S_{BH}$ is the BHAF, it is clear that logarithmic terms in horizon area must arise. This limiting procedure is considered to be a quitessentially quantum gravity effect, as argued in \cite{pm2025}. 

Now, HAT-validating GW data analyses from BBHC observations choose unequivocally a {\it negative algebraic sign} of the correction coefficient $s_0$, for {\it absolute consistency}, in the same spirit that the classical HAT imposes restrictions on physically permitted black hole dynamics. For BBHC, with two not too dissimilar isolated black holes spiralling around each other and eventually merging into one larger black hole post ringdown, with emission of gravitational waves, the HAT does not categorically quantify how much larger the final stationary configuration horizon area has to be, in respect of the sum of the inspiralling black hole horizon areas. The difference between the post-merger remnant black hole and inspiral black hole horizon areas only need obey a positive algebraic sign. In this spirit, we treat the result of the analyses of \cite{teu20}-\cite{bad22} validating the HAT for BBHC observations by LIGO-VIRGO-KAGRA as providing the constraint of an algebraic sign on the coefficient of the logarithmic corrections to the BHAF computed by various means \cite{pm2025}. Violation of the criterion of {\it absolute} consistency does not make the result of the correction computation to BHAF inconsistent with GW observational validation of the HAT, because {\it relative} consistency is still possible. My adherence to the absolute consistency criterion is because, while embodying the spirit of the HAT in its entirety, it does provide in addition an opportunity of constraining, on the basis of GW observations, aspects of quantum gravity formulations. Such constraining of quantum gravity formulations on the basis of GW observational results might be deemed impossible otherwise. This makes it worthwhile to explore the implications of this criterion for logarithmic corrections to the BHAF  in the case of BBHC, calculated on the basis of various quantum gravity models.    

Ever since Bekenstein's pioneering proposal \cite{bek73} of the BHAF (supplemented by Hawking's seminal paper \cite{haw75} on black hole radiance), deriving this formula from a set of `first' principles has been the goal of a large part of the literature on black hole thermodynamics. While Bekenstein's proposal mentioned that the origin of the microstates leading up to the BHAF must be {\it quantum} gravitational in nature, the absence (even until today) of a complete theory of quantum gravity has led to some confusion as to whether the BHAF encapsulates microstates of the quantum spacetime itself, or of the {\it environment}, i.e., of matter fields in the vicinity of a classical black hole spacetime. This dilemma has spawned two distinct classes of approaches to the problem, namely, {\it background independent} studies of quantum spacetime fluctuations without and with quantum matter, and the study of quantum matter field fluctuations in a {\it non-dynamical} classical black hole background. Models like Loop Quantum Gravity (LQG), Causal Dynamical Triangulations (CDT), Causal Sets (CST) and so on, fall in the first category of quantum gravity formulations. These formulations can be clubbed into the term `quantum geometry' and are inherently non-perturbative in nature. The latter class includes the Brick-Wall model based on the Gibbons-Hawking Euclidean Quantum Gravity formlalism, followed by various approaches employing ideas of {\it entanglement entropy} (earlier called `geometric' entropy) \cite{cal1994}-\cite{sol2020}  and are inherently {\it perturbative} in nature, around a classical black hole spacetime background. 

The gravity-matter entanglement idea (see, e.g., \cite{bskay2022}-\cite{bskay2023}) claims that for black holes in equilibrium, the quantum gravitational degrees of freedom are already included in the entanglement of matter states with graviton states within perturbation theory. This procedure cannot yield the BHAF for at least two reasons \cite{sol2011} : first of all, the BHAF has no allusions to any matter fluctuation quantum number, being a function entirely of the black hole horizon area. Secondly, the entanglement entropy at any order of perturbation theory, must include an ultraviolet cutoff, whereas the BHAF is a finite quantity.  In fact, the gravity-matter entanglement approach misses out the {\it non-perturbative} pure quantum geometry states around {\it zero background}; these are not represented by perturbative graviton states around a fixed classical black hole spacetime. On the other hand, LQG as applied to Isolated Black Hole Horizons \cite{abh1998}-\cite{abh2000} is based on such non-perturbative horizon states, as quantum Chern-Simons states, and account for an ab initio derivation of the BHAF, as well as logarithmic corrections thereof \cite{km98}-\cite{abhi-pm14}. For macroscopic black holes, no ultraviolet issues plague this approach. Additional logarithmic corrections from entanglement entropy do indeed arise; these carry explicit signatures of quantum matter fluctuations (in terms of spin and number of species for each spin); furthermore, they require ultraviolet renormalization for their proper incorporation \cite{sol2011}-{\cite{sol2020} as genuine logarithmic corrections.    

In this paper, we investigate the implications of the {\it totality} of all possible logarithmic corrections to the BHAF for the entropy of a black hole, due both to non-perturbative quantum spacetime fluctuations around vanishing backgrounds, as well as, due to perturbative quantum matter and graviton fluctuations in non-dynamical black hole backgrounds. An inherently non-perturbative scheme as in various quantum geometry approaches, like say LQG, is difficult to use to extract perturbative information due to matter fluctuations around a non-dynamical black hole background. The difficulty is comparable to doing perturbative calculations at high energies in quantum chromodynamics (QCD), within the Lattice Gauge Theory methodology which is inherently geared towards gleaning non-pertubative low energy information regarding the strong interactions in elementary particle physics. However, Lattice QCD alone cannot yet produce parton distribution functions which are phenomenologically relevant, because Lattice QCD operators are subject to perturbative renormalization effects (see, e.g., ref. \cite{mcon23}). When extrapolated to the issue of computing logarithmic corrections to the BHAF, difficulties in combining background-independent, non-perturbative LQG results to perturbative, background-dependent. entanglement entropy results are of a similar character. Perturbative renormalization of LQG operators employed to compute logarithmic corrections to the BHAF, due to matter and graviton fluctuations, are not available for all length scales. 

Yet, if the issue of absolute consistency with GW observational results is to be taken seriously, a model of combining the perturbative and non-perturbative approaches needs to be proposed, so that the net effect of observational results on logarithmic corrections to the BHAF can be probed. Keeping possible caveats in view, we {\it propose} that the non-perturbative coefficient $s_0^{lqg}$ be algebraically added on to the perturbative entanglement entropy result $s_0^{ent,1}$. As we shall see later, this proposal has interesting implications for elementary particle cosmology. There is a further commonality between the LQG and entanglement approaches : the role of conical singularities in arriving at the final result for the logarithmic corrections. In the LQG case, these conical singularities result from the punctures produced by spin network edges on the horizon 2-sphere. They are resolved by the minimum area result of LQG : area eigenvalues are bounded from below. In the case of entanglement entropy based on Euclidean Quantum Gravity around a classical black hole background, the matter and graviton fluctuations need to be periodic in Euclidean time, with the period identified with the inverse temperature. This periodicity is the source of the conical singularity which is resolved by setting the inverse temperature to the inverse Hawking temperature \cite{cal1994}. In the Replica method, the entanglement entropy is evaluated from the partition function by a derivative with respect to the deficit angle of the conical singularity, for each Riemann sheet at the branch point and then summing over all contributions.           

Section 2 in mainly a review. I recapitulate briefly both the basics of HAT, applicable to BBHC, and the results of recent analyses of GW data for BBHC which validate the HAT to a remarkable degree of accuracy. In the same section, the BHAF and the Generalized Second Law (GSL) are reviewed, as appropriate to BBHC. Section 3 is where the main contributions are exhibited. It addresses the issue of logarithmic corrections to the BHAF. Subsection A discusses some generalities, briefly summarized, especially from the recent ref. \cite{pm2025}. Discussions of the methodologies and results of LQG and the Entanglement Entropy approaches, follow. We then seek to establish absolute consistency of the logarithmic correction results of each approach individually, and also in combination with physical justification. The section ends with constraints on the spectrum of quantum matter fluctuations that follow from the absolute consistency criterion, and their implications for particle cosmology. I conclude in section 4 with some remarks.          
      
\section{Recap of the Basic Framework}

\subsection{Hawking Area Theorem and GW Data from BBHC Observations} 

The black hole spacetime ${\cal B} = {\cal M}  - J^-({\cal I}^+)$, with $ \partial {\cal B} = h~,~h \cap {\Sigma}_{1,2} \sim S^2 (topol)$, where ${\cal M}$ is the entire spacetime, and $J^-({\cal I}^+)$ is the past of all events which reach ${\cal I}^+$. For isolated black holes, $A(h \cap \Sigma_1) = A(h \cap \Sigma_2) \equiv A_h$, where ${\Sigma}_{1,2}$ are spatial foliations, with $\Sigma_1$ corresponding to a time slice {\it earlier} than $\Sigma_2$. This implies that the horizon area of an isolated black hole is the same on all spatial foliations, i.e., it is {\it conserved}. However, for accreting black holes, $A(h \cap \Sigma_1) < A(h \cap \Sigma_2) \Rightarrow A_{h,fin} > A_{h,ini}$, i.e., the horizon area must increase. In summary, the horizon area can {\it never decrease} in any physical process ! This is the Hawking Area Theorem (HAT) \cite{haw71}. An immediate consequence of the HAT is that a large black hole can never disintegrate into two smaller black holes, while two inspiralling black holes may orbit each other and coalesce into a larger remnant black hole with emission of gravitational waves, the binary black hole coalescence (BBHC). In a BBHC, therefore, we have the inequality 
\begin{eqnarray}
A_{h,rem} &>& A_{h,1} + A_{h,2} \Rightarrow (\Delta A_h/A_{h,i}) > 0 \nonumber \\
\Delta A_h & \equiv & A_{h,rem} - (A_{h,1} + A_{h,2}) \nonumber \\
A_{h,i} &=& A_{h,1} + A_{h,2}\label{hat}
\end{eqnarray}

The LVK consortium has made a series of observations of gravitational wave emission events which can be considered as BBHCs. Detailed analyses of the GW waveforms for the inspiral, merger and ringdown phases have been performed by several groups for GW150914 and GW170914. In ref. \cite{teu20}, comprehensive analyses of data from GW150914 have been presented, with details of the probability density of events as a function of $\Delta A/A_i$. Choices have been made with respect to times of the onset of ringdown and the end of the merger phase, as also, if frequencies beyond the fundamental mode for the GW waveforms have to be included. In ref. \cite{bad22}, a somewhat technically different approach has been adopted to analyse GW data from BBHCs observed for GW150914 and GW170914, with disparate choices about the onset of ringdown and end of the merger  phase. But the overall consensus is of course that the HAT is validated with high probability, i.e., $\Delta A_h/A_{h,i} > 0$. This is the key observational result that I shall use in what follows. Notice that this result is solely about the algebraic sign of the relative change in horizon area $\Delta A_h/A_{h,i}$; the observed magnitude of this quantity is not of relevance.  

\subsection{The BHAF}

Isolated systems in ordinary thermodynamics are characterized by their conserved energy ${\cal E}$, their volume and other extensive quantities. There then exists the energy function $S({\cal E}, ...)$ which is non-negative and additive for isolated compound systems. When such systems lose their isolation and interact, with exchange of energy, from an initial value ${\cal E}_i$ to a final value ${\cal E}_f$, returning to equilibrium at the end of the interaction, the entropy obeys the inequality $S({\cal E}_f) > S({\cal E}_i)$. 

For vacuum black hole solutions of general relativity, the bulk energy vanishes as a Hamiltonian constraint ${\cal E} \approx 0$. On the other hand, the horizon area $A_h$ is a conserved quantity. So following Bekenstein \cite{bek73}, one can define a black hole entropy $S_{bh} = S_{bh}(A_h)$ which is non-negative and additive. When the black hole accretes matter from a neighbouring large star, it's horizon area changes, say from $A_{h,i}$ to $A_{h,f} > A_{h,i}$ (by the HAT), the black hole entropy must obey the inequality $S_{bh}(A_{h,f}) > S_{bh}(A_{h,i})$. During accretion of matter, the matter entropy also changes, so that one has the Generalized Second Law (GSL) in a universe with black holes present alongwith ordinary matter,
\begin{eqnarray}
S_{bh}(A_{h,f}) + S_{mat,f} > S_{bh}(A_{h,i}) + S_{mat,i} \label{accr}
\end{eqnarray}  
For observed BBHC events, since there is not much data yet available on accretion of the inspiralling pair, the GSL takes the form 
\begin{eqnarray}
S_{bhr}(A_{hr}) + S_{GW} > S_{bh1}(A_{h1}) + S_{bh2}(A_{h,2}). ~\label{gsl}
\end{eqnarray}
where $S_{GW}$ is the entropy carried by the GW emission. This quantity has been estimated \cite{mr21} and found to be negligibly small compared to all black hole entropies; we shall therefore ignore this contribution in what follows.

Assuming that the horizon area $A_h$ is an {\it extensive} quantity in black hole thermodynamics, Bekenstein \cite{bek73} made the proposal that 
\begin{eqnarray}
S_{bh}(A_h) = \frac{A_h}{A_{Pl}} \equiv S_{BH}(A_h)~,~ A_{Pl} = \frac{4 G \hbar}{c^3} \label{bhaf}
\end{eqnarray}
We have subsumed the factor of $4$ (due to Hawking \cite{haw75}) into the definition of the Planck area $A_{Pl}$, whose choice as the divisor in Bekenstein's proposal above is argued on the basis of {\it universality}. But it also has the implication, emphasized in ref. \cite{bek73} that the physical origin of black hole entropy must ultimately be the `atoms' of gravity, i.e., the degrees of freedom of a theory of {\it quantum gravity}. 

\section{Logarithmic Corrections to the BHAF}

\subsection{General Features and Absolute Consistency with GW Observations}

It is clear that together with eqn. (\ref{gsl}), the observed GW data from BBHC validating eqn (\ref{hat}), implies that $(\Delta S_{BH})_{obs} \equiv S_{BHr} - (S_{BH1} + S_{BH2})_{obs} > 0$. However, the very fact that an {\it ab initio} calculation of black hole entropy must necessarily involve a quantum gravity proposal, raises the possibility that theoretical computations of black hole entropy within that proposal (or any other) may throw up {\it quantum corrections} to the BHAF (\ref{bhaf}). As argued in ref. \cite{pm2025}, for astrophysically relevant black holes ($S_{BH} >> 1$), black hole entropy $S_{bh} = S_{bh}(S_{BH}, {\bf Q})$, where ${\bf Q}$ are a bunch of `charges' or `hair' characterizing a modification of general relavity, either quantal or classical. A simple assumption is that these corrections to the BHAF are {\it additive}, i.e., one may write \cite{pm2025}
\begin{eqnarray}
S_{bh} &=& S_{BH}(A_h) + s_{bh}(S_{BH}, {\bf Q}) \nonumber \\
s_{bh} & \simeq & s_0({\bf Q}) \log S_{BH} + s_1 ({\bf Q}) S_{BH}^{-1} + \cdots . \label{cor} 
\end{eqnarray}    
Substitution of eqn. (\ref{cor}) into (\ref{gsl}) leads to the result \cite{pm2025}
\begin{eqnarray}
- \Delta s_{bh} < (\Delta S_{BH})_{obs}~,~\Delta s_{bh} \equiv s_{bhr} - (s_{bh1} + s_{bh2}). \label{corbd} 
\end{eqnarray}

The GW data from BBHC asserts $(\Delta S_{BH})_{obs} > 0$; correspondingly, we define the possibility from eqn (\ref{corbd}) that $\Delta s_{bh} > 0$ as {\it absolute consistency} of the theoretical computation of the corrections $s_{bh}$ with observational GW data corresponding to BBHC. We shall adhere to this definition henceforth for the remainder of this paper, modulo the caveat mentioned in the Introduction.  

Restricting our considerations to solely the $\log S_{BH}$ corrections, it has been shown in ref.\cite{pm2025}, with some rather general assumptions, absolute consistency with GW data on BBHC requires that $s_0 = -|s_0|$, i.e., the corrected black hole entropy must be lower than the BHAF value. How does this result compare with computations based on quantum gravity proposals ? 

\subsection{The LQG Approach}
  
As discussed in ref. \cite{pm2024}, the Hilbert space in this approach is ${\cal H} = {\cal H}_{\cal M} \otimes {\cal H}_{\partial \cal M}$, where ${\cal M}$ is bulk spacetime (including bulk matter), and its boundary is $\partial {\cal M}$ which is identified with the horizon $h$ of an isolated black hole as an {\it inner} horizon of spacetime \cite{abh1998} with no radiation or matter  on it. Thus, any state $|\Psi \rangle \in {\cal H}$ can be written as 
\begin{eqnarray}
|\Psi \rangle = \sum_{{\cal M},h} D_{{\cal M} h} |\psi \rangle_{\cal M} \otimes |\psi \rangle_h \label{hil}
\end{eqnarray}
where the coefficient $D_{{\cal M} h}$ is {\it not} assumed to be diagonal. The inclusion of matter in bulk spacetime background is done through having ${\cal H}_{\cal M} = {\cal H}_g \otimes {\cal H}_M$, so  that bulk states are given as 
\begin{eqnarray}
|\psi \rangle_{\cal M} = \sum_{g, M} C_{gM} |\psi \rangle_g \otimes |\psi \rangle_M \label{hblk}
\end{eqnarray}
where, once again, the coefficients $C_{gM}$ are not assumed to be diagonal, implying leaving open the possibility of gravity-matter entanglement. The bulk Hamiltonian 
\begin{eqnarray}
{\hat H}_{\cal M} = {\hat H}_g \otimes {\bf I} + {\bf I} \otimes {\hat H}_{M} \label{blkh}
\end{eqnarray}
annihilates physical states in the bulk : 
\begin{eqnarray}
{\hat H}_{\cal M} | \psi \rangle_{\cal M} = 0 \label{hcon}
\end{eqnarray}
The full Hamiltonian 
\begin{eqnarray}
{\hat H} = {\hat H}_{\cal M} \otimes {\bf I} + {\bf I} \otimes {\hat H}_h
\end{eqnarray}
has the expectation value from eqn. (\ref{hil}) \cite{pm2024}
\begin{eqnarray}
\langle \Psi | {\hat H} | \Psi \rangle &=& \sum_h {}_h\langle \psi'| {\hat H}_h |\psi' \rangle_h \label{expc} \\
|\psi' \rangle_h & \equiv & \sum_{\cal M} | D_{{\cal M} h} |~ || |\psi \rangle_{\cal M} || ~| \psi \rangle_{h}  
\end{eqnarray}
Thus, the holographic character of the expectation value of the full Hamiltonian is a direct consequence of the bulk Hamiltonian constraint eqn. (\ref{hcon}). The horizon states obey the quantum Chern-Simons equation of motion on the horizon $h$ \cite{abh1998}
\begin{eqnarray}
\frac{A_h}{2\pi A_{Pl}} {\hat F}^{IJ}_{ab} | \psi \rangle_h &=& - \sum_{ \{ p \} } {a}_{ab}(p) {\hat {\cal J}^{IJ}(p) } | \psi \rangle_h \label{cseom} \\
{\hat {\cal J}^{IJ} }(p) &=& {\hat {\cal J}^{IJ} }_g + {\hat {\cal J}^{IJ} }_M
\end{eqnarray} 
In eqn (\ref{cseom}), $a_{ab}(p)$ corresponds to the area function around the conical singularity at the puncture $p$ of the two dimensional spacelike surface $h \cap \Sigma$ where $\Sigma$ is a spatial hypersurface foliating the null horizon in a 2-sphere. As already mentioned in the Introduction, the lower bound on the value of this area quantum, derived as an exact result in LQG,  resolves the conical singulalrity at the puncture. The spin operators ${\hat {\cal J}^{IJ}}(p)$ correspond to the spin of the bulk spin network edge creating the conical singularity at the puncture $p$ on  $h \cap \Sigma$ \cite{pm2024}. The number of horizon microstates $\Omega =  \dim \{ |\psi \rangle_h \} = \dim {\cal H}_h$ and has been calculated \cite{km2000}. The quantum-corrected black hole entropy is given, for astrophysical or macroscopically large area black holes, by 
\begin{eqnarray}
S_{bh} = \log \Omega = S_{BH} - \frac32 \log S_{BH} + {\cal O}(S_{BH}^{-1}) . \label{kms}
\end{eqnarray} 
Thus, we find that $s_0^{lqg}=-3/2$. The most important aspect of this formula is its absolute consistency with GW observational results, as outlined in the last subsection. The other noticeable feature is the complete finiteness of this result, without any need of any ultraviolet renormalization. Finally, the fact that the result for quantum black hole entropy is only a function of the BHAF for black hole entropy, underlines its precise origin in quantum spacetime geometry.

\subsection{The Entanglment Entropy Approach}

\subsubsection{Summary of the approach}

So, the issue, vis-a-vis our notion of absolute consistency with GW observations, reduces to examination of results from the perturbative `geometric' (or entanglement) entropy calculation. This has been reviewed by Solodukhin \cite{sol2020} about five years ago, whose results we quote in this paper. Although perturbative computations have certainly gone on, it is not clear that ultraviolet divergence issues related to perturbative Euclidean quantum gravity have all been resolved. Hence, we restrict our attention to the least ambiguous one-loop results discussed in ref.\cite{sol2020}.

The Euclidean partition function for matter field and graviton fluctuations ($\Phi_m, {\tilde g}$) around a spherically symmetric (Schwarzschild) classical black hole background, characterized by a metric $g_{bh}$, is given by
\begin{eqnarray}
Z[g_{bh}] = \int {\cal D}{\tilde g} {\cal D} \Phi_m \exp - {\cal I}(g_{bh}, {\tilde g}, \Phi_m) \label{epf}
\end{eqnarray} 
For finite (inverse) temperature $\beta$, require all fields in $Z$ to be periodic under $\tau \rightarrow \tau + \beta$. This implies that $ Z[g_{bh}, \beta]$ develops conical singularities for arbitrary $\beta$ with deficit angle $\delta_{\beta} = 2\pi (1-\beta/\beta_H)$ at $h$, where $\beta_H^{-1} \rightarrow$ Hawking temperature for $g_{bh}$ on $h$. This, leads to the interpretation \cite{cal1994} that a black hole is in equilbrium only at a temperature equal to its Hawking temperature (surface gravity at its horizon). Now, using the definition of the equilibrium entropy in terms of the canonical partition function, one obtains \cite{sol2011}
\begin{eqnarray}
S_{bh}^{ent} = \left( 2\pi \frac{d}{d \delta_{\beta}} + 1 \right) Z(\delta_{\beta})|_{\delta_{\beta}=0} ~\label{geos}
\end{eqnarray} 

\subsubsection{One Loop Results of the Logarithmic Correction Computation}

Upon computing the one loop corrections to the BHAF, the result, as expected, turn out to be ultraviolet divergent; to ameliorate this, one needs to {\it renormalize} the Planck area $A_{Pl}$, or equivalently, the Newtonian constant $G$, to obtain a finite logarithmic correction. The net upshot is a formula for the one loop correction to the BHAF, expressed in terms of the spectrum of elementary particle excitations in the black hole background, in terms of the spin of the excitations, and the number of species of each spin \cite{sol2011}
\begin{eqnarray}
s_{bh} &=& s_0^{ent,1} \log S_{BH} + \cdots \\
s_0^{ent,1} &=& \sum_{j=0}^2 N_j a_j, ~j=0^{\pm}, \frac12, 1, \frac32, 2. \label{sgeo} 
\end{eqnarray}
In eqn(\ref{sgeo}), $N_j$ is the number of species of spin $j$ matter (or graviton) fluctuations, and $a_j$ the one loop perturbative contribution of matter fluctuations of that spin around a classical black hole spacetime background. The coefficients $a_j$ have actually been computed \cite{sol2020} explicitly around a four dimensional Schwarzschild spacetime background, yielding the result
\begin{eqnarray}
s_0^{ent,1} &=& \frac{1}{45} [ N_{0^+} + \frac72 N_{1/2} -13 N_1 \nonumber \\
&+& 91 N_{0^-} - \frac{233}{4} N_{3/2} + 212 N_2 ] ~\label{sgeo1}
\end{eqnarray}
The first line of eqn (\ref{sgeo1}) includes observed matter fields appropriate to the spectrum of the Standard Electroweak-QCD Theory of elementary particle physics. The second line includes matter fluctuations which have been observed neither at the LHC (or any other elementary particle physics experiments), nor in astrophysical or cosmological observations, and constitute what is commonly called Beyond Standard Model (BSM) particles. 

It is fairly clear that if we impose the criterion of absolute consistency on the coefficient $s_0^{ent,1}$, ignoring the LQG contribution,  i.e., it must have a negative sign by itself, then restrictions will emerge on $N_j$. The one loop result for gravity-matter entanglement entropy quoted above counts only physical degrees of freedom involved in gravity-matter entanglement, irrespective of their masses or energies. Energy thresholds typically show up at higher orders of perturbation theory. So it is independent of any proposed particular model of BSM physics. For the Standard EW-QCD spectrum, $N_{0^+} = 1, N_{1/2} = 24, N_1 = 12, N_{0^-} = N_{3/2} = N_2 = 0$. In this case, obviously absolute consistency with the GW data is guaranteed. On the other hand, the balance is tilted towards relative consistency from absolute consistency with GW data, if we have $N_{0^-}=1=N_2$ unless we compensate it with sufficient number of spin $3/2$ gravitinos, $N_{3/2}=3$. If gravitinos are never observed, then the coexistence of scalar axions and spin 2 gravitons is in doubt, according to this prediction.           

\subsection{Combining the LQG and Entanglement Entropy Logarithmic Correction Results}

As discussed in the Introduction, the basic difficulty in combining the non-perturbative logarithmic corrections with the perturbative results of the last subsection is that it is not clear that they are at the same energy scales. Unlike the case of QCD which is a {\it perturbatively ultraviolet renormalizable} gauge theory in a flat spacetime background, it is hard to see how to renormalization group improve the perturbative entanglement entropy result to match the non-perturbative LQG result, so that the former can be deemed as a {\it renormalization} of the latter. As such, even a formula like eqn (\ref{sgeo}) emerges only {\it after} an ultraviolet renormalization of the Planck area (equivalently the Newton gravitation constant) \cite{sol2020}. A further degree of complication arises because of background dependence : the LQG computation is background-free, while the entanglement entropy is around a classical Schwarzschild spacetime background. 

Despite these caveats, there are clear similarities in the two seemingly disparate approaches to quantum black holes. The actual counting in LQG of isolated black hole horizon microstates employs properties of a two dimensional conformal field theory, namely the number of conformal blocks of an $SU(2)$ WZW theory on the 2-sphere foliation of the horizon, with point sources at the conical singularities present at the punctures made by spin network edges. The conical singularities are smeared out by the existence of a lower bound on the area eigenvalues on the quantum isolated horizon in LQG. Similarly, the entanglement entropy approach employs conical singularities in Euclidean time on the 2-sphere foliation of the horizon, due to periodicity at finite temperature, resolved by the Hawking temperature. The Euclidean partition function is a function of the deficit angle at the conical singularity for each replica of the horizon, eventually to be summed over all copies to yield the entanglement entropy, modulo ultraviolet renormalization. In this sense, the perturbative result $s_0^{ent,1}$ can be deemed to be a {\it renormalized version} of the LQG coefficient $s_0^{lqg}$ of the logarithmic correction to the BHAF, leading to the result   
\begin{eqnarray}
s_0 &=& s_0^{lqg} + s_0^{ent,1} . \label{hyb}
\end{eqnarray}
Of course, this formula is subject to improvement by incorporation of higher order pertubative results for the entanglement entropy. But here we do not discuss these higher loop effects. As explained in an earlier subsection, absolute consistency with GW data analyses imposes the restriction $s_0 < 0$. As already mentioned, even a formula like (\ref{hyb}) for the net coefficient of the logarithmic corrections to BHAF, with the absolute consistency restriction, will impose constraints on the number of species $N_j$ of spin $j$ matter and graviton fluctuations around the Schwarzschild background. These constraints are easy to determine using (\ref{hyb}), together with eqn.s (\ref{kms}) and (\ref{sgeo1}).  

\subsection{Implications}

Evaluating the numerical result for $s_0$ from eqn (\ref{hyb}), we get, in the case of exclusively the spectrum of the Standard EW-QCD Theory, the result $s_0 = -277/90 < 0$, implying absolute consistency with GW observational results. We continue to set $N_{3/2}=0$ because gravitinos are gauge fermions of supergravity, and there is no observational evidence yet of supersymmetry, and hence of supergravity. With additional contributions from $N_{0^-} = 1 = N_2$, we get $s_0 = 143/30 > 0$. Thus, coexistence of a single species of axions and a single species of spin $2$ gravitons exhibits a {\it tension} with our criterion of absolute consistency. I hasten to reiterate a disclaimer : this is not to say that such a coexistence is {\it inconsistent} with GW results from BBHC observations. But, in my opinion, it is important to recognize the difference between absolute and relative consistency, especially vis-a-vis the GW observations for BBHC. Relative consistency is in some sense {\it trivial} because of the huge separation of scales between quantum gravity computational results and GW observational results. On the other hand, absolute consistency is {\it nontrivial} because it involves algebraic signs exactly like the HAT does. Also, despite our best efforts, BSM particles - axions or gravitons - have not yet been observed in any experiment or astrophysical or cosmological observation. Even an indirect theoretical rationale for the existence or otherwise of these particles is of substantive current relevance. This is of course not to claim that these particles will never be observed, just like the issue of existence of magnetic monopoles or fractionally charged particles. If the said coexistence of axions and gravitons is actually observed in the foreseeable future, then our criterion of absolute consistency with GW data will be falsified immediately. Therein lies its primacy over relative consistency.

\section{Conclusions}

The constraints on the BSM spectrum of particles may be thought of as constraints on some Dark Matter candidates, especially axions. Even in the absence of gravitons and gravitinos, the absolute consistency criterion rules out the existence of more than a single species of axions. Now, as far as I know, there are no genuine constraints, astrophysical or cosmological, on the existence and characteristics of axionic particles as candidate Dark Matter particles, despite the proliferation of BSM particle physics models including axions.   

On this last issue, we may reiterate that the one loop contribution $s^{ent,1}_0$ due to quantum matter field fluctuations to entanglement entropy, quoted above, does not depend on the energy at which these fluctuations are observed. So long as they are {\it physical} degrees of freedom with gravitational interaction (and thus `entangled' with graviton states), their masses (energies) play no role at the one loop computation of the entanglement entropy. BSM models which may involve fundamental fields at very high energy beyond accessibility at the present time, shall also contribute to this entropy calculation, since they interact with gravity at {\it all} energies. The black hole entropy, in this definition, merely {\it counts} the fundamental field degrees of freedom entangled with the black hole horizon, and hence has a universal character, although the details do depend on the spin-parity and number of species of these fields.    

The restriction, albeit indirect, on the spectrum of field fluctuations which lead to a net composite logarithmic correction to the BHAF for black hole entropy, which is absolutely consistent with GW observations, is in some ways reminiscent of the restriction on the particle spectrum in an {\it asymptotically free} gauge theory of particle interactions (excluding gravitons). As briefly discussed in ref. \cite{sol2020}, there may be a theoretical link between the geometric or entanglement entropy approach to black hole entropy corrections, and the renormalization group equations expressing the behaviour of gauge field theory running couplings under momentum scaling. This relationship is probably known to some extent, although a full elucidation will be highly desirable.

On the question of observability of spin $2$ gravitons at sub-Planckian energies, there are powerful arguments due to Dyson \cite{dyson13} that this may not be possible. An easy way to paraphrase this is to follow 't Hooft and postulate two energy-dependent, {\it dimensionless} gravitational coupling parameters, from Newton's gravitational constant $G : g_s \equiv Gs, g_t \equiv Gt$, where $s,t$ are Mandelstam kinematical invariants. It is fairly obvious that graviton scattering cross-sections at sub-Planckian (e.g., LHC) energies are going to be negligibly small. In other words, gravitons may well exist, but their observation might entail Planckian energies at which the two dimensionless coupling constants defined above are no longer perturbative. 

Finally, we point out that the very existence of Dark Matter as necessary for a consistent understanding of the laws of Nature, has faced some criticism for sometime, by some gravitational physicists \cite{lud2021}-\cite{wilt2024}, claiming to be able to reproduce the galactic rotation curves by appropriately modelling the galactic rotation and the geometry it reproduces, directly from general relativity with simply ordinary visible matter. So, some aspects of the arguments leading to the existence of Dark Matter is already under debate, implying that such existence cannot be taken for granted, especially in light of non-observation of any Dark Matter particle to date. The issue of weak gravitational lensing of photons from galaxies or clusters of galaxies, used as the other argument in favour of the existence of Dark Matter, ought to be revisited also for corroboration.  

\section{Acknowledgement}

We gratefully acknowledge illuminating discussions with Benito Aubrey, Vitor Cardoso, Paolo di Vecchia, Sayan Kar, Bernard Kay, Jorma Louko, Debasish Majumdar and Sandipan Sengupta.

\end{document}